# "Quantum" key distribution using weak classical light waves

**Sergey A. Rashkovskiy**

*Ishlinsky Institute for Problems in Mechanics of the Russian Academy of Sciences, Vernadskogo Ave., 101/1, Moscow, 119526, Russia*

*Tomsk State University, 36 Lenina Avenue, Tomsk, 634050, Russia*

*E-mail: rash@ipmnet.ru, Tel. +7 906 0318854*

**Abstract** The detection of very weak classical electromagnetic (light) waves by classical macroscopic device is discussed. It is shown that the results of such detection can be interpreted as a manifestation of the quantum properties of radiation, although in reality they are related to the peculiarities of the interaction of weak classical electromagnetic waves with discrete atoms. We show that the "quantum" key distribution protocol can be realized using very weak classical light waves and avalanche detectors, and it possesses all the properties of the quantum cryptographic protocol E91 which based on entangled photons.

## 1 Introduction

The entanglement phenomenon is one of the most mysterious and intriguing predictions of quantum mechanics. It manifests in a strong correlation of the behavior of quantum objects, even when they are separated by a large distance. One of the practical applications of the entanglement, actively discussed in the literature, is the quantum key distribution. The basis for this is the E91 protocol [1], which uses Bell's inequality. The most important property of this protocol is the impossibility to eavesdrop the key imperceptibly. This is due to the indivisibility of the photon and to the fundamental feature of the entangled state of two photons: after interception (detection) of one of a pair of entangled photons, entanglement destroys; this means that in principle it is impossible to create an exact copy of an intercepted photon which would be in an entangled state with a second (non-intercepted) photon (the no-cloning theorem).

It was shown in [2-10] that many quantum phenomena can be explained in a natural way and described in the framework of classical field theory without quantization of radiation. In particular, all the features of the EPRB Gedankenexperiment with entangled photons can be reproduced using very weak classical light waves [10], and described in natural way within the framework of classical wave optics if we take into account the features of the interaction of a classical electromagnetic wave with an atom. In this case, paradoxes such as "wave-particle duality" of light and "spooky action at a distance", which are an mandatory attribute of photonic interpretation, do not arise.

As shown in [10], what is called "entanglement of photons" in quantum mechanics, for weak classical light waves is explained by their correlation.

The results obtained in [10], inevitably lead us to the question: can the classical "entangled" light waves be used for quantum computing and quantum cryptography? Is it possible to use the classical correlated light beams taking into account their specific character of interaction with the detectors to construct the computational algorithms, similar to "quantum" algorithms?

In this paper, being based on the results of work [10], we describe a key distribution protocol, similar to the E91 "quantum" key distribution protocol [1], but which uses only weak classical light waves and a feature of their interaction with the atoms of the detector. The process is classical at all stages, because of this we do not encounter paradoxes such as "wave-particle duality" of light and "spooky action at a distance".

## 2 Features of the detection of the very weak classical electromagnetic (light) waves

First of all, we note that all the electromagnetic (light) waves considered below are classical and fully described by classical electrodynamics or even classical wave optics.

Let us determine in what cases the action of the electromagnetic (light) wave can be considered weak.

To do this, let us consider a monochromatic wave that acts on an elementary detector, which can be an atom, a molecule, or a photosensitive crystal of a measuring device.

Solution of the Schrödinger equation allows calculating the probability of excitation of an atom of the detector by classical electromagnetic (light) wave for time $\Delta t$ (Fermi's golden rule)

$$w\Delta t = bI\Delta t \qquad (1)$$

where $w$ is the probability of excitation of atoms per unit time; $I$ is the intensity of the classical light wave at the location of the atom; and $b$ is a constant which for not very intense beam does not depend on the intensity of incident light.

As shown in [2-4], the probability of excitation of an atom during a time $\Delta t$ is

$$P(t) = 1 - \exp(-w\Delta t) \qquad (2)$$

Taking into account (1), one obtains

$$P(t) = 1 - \exp(-bI\Delta t) \qquad (3)$$

It follows that when

$$bI\Delta t \gg 1 \qquad (4)$$

the excitation of detector atoms will occur with a probability close to unity, and the process of detecting a light wave in this case will be classical and deterministic.

For this reason, impact satisfying condition (4) will be considered strong (classical), while impact that does not satisfy this condition will be considered weak. Obviously, at a weak impact, the detector will operate with a certain probability not equal to unity, and the detection process will be random.

Taking into account that $I \sim \mathbf{E}^2$, relation (4) can be rewritten in the form

$$b\mathbf{E}^2 \Delta t \gg 1 \qquad (5)$$

Let us consider qualitatively the process of detecting classical electromagnetic waves by some classical macroscopic (i.e., consisting of a large number of atoms) measuring device.

Let us first assume that the action of a light wave on the detector is strong and satisfies condition (4). In this case, under the action of this wave, a very large ($N \gg 1$) number of atoms of a macroscopic measuring device will be excited simultaneously, which will cause macroscopic electric currents in the device, which the device will detect. The greater the intensity (electric field strength) of the light (electromagnetic) wave, the more electric current will be induced by the wave in the measuring device, which will be continuously reflected in the instrument readings. Thus, using such a device, it is possible to continuously measure various characteristics (intensity, polarization, phase, etc.) of the wave (4). Let us begin to decrease the intensity of the incident wave. In this case, the number of simultaneously excited atoms of the measuring device will decrease, and we will observe fluctuations in the readings of the measuring device, which will increase as inequality (4) is violated. These fluctuations are of a classical (non-quantum) nature and are associated with the discrete (atomic) structure of the detector, with thermal fluctuations, as well as with fluctuations in the intensity of the incident classical light wave that always occur, and increase with decreasing wave intensity. This is due to the nature of the emission of the electromagnetic wave, the sources of which are individual atoms of the emitter: when the intensity of the wave is large, this means that the number of simultaneously emitting atoms was large and their fluctuations compensate each other; at a low intensity of the light wave, the number of simultaneously emitting atoms was small and the random nature of the emission of individual atoms will be manifested more strongly in the form of fluctuations in the wave parameters (amplitudes of electric field strength, phase and polarization). If the inequality (4) is strongly violated, only small groups of atoms or even individual atoms will be excited at the same time in the measuring device. In this case, we cannot continuously measure the parameters of the incident wave: the fluctuations in the readings of the device will be so great that we cannot separate the true signal from the noise. In other words, light (electromagnetic) waves, for which condition (4) is not satisfied, cannot be continuously measured by classical macroscopic devices. Such waves can be detected only by avalanche detectors, i.e. detectors in which the excitation of one atom or a small group of atoms induces an electron avalanche that manifests itself in the form of a short pulsed electric current (click of detector). It is obvious that the avalanche detector does not allow measuring continuous parameters (electric field strength, phase, polarization) of a very weak classical electromagnetic (light) wave, since the only reaction of the avalanche detector to the action of such a wave is a discrete click (an electric current pulse). Thus, the avalanche detector allows detecting only the very fact of the excitation of an atom (or group of atoms) under the action of a very weak classical light (electromagnetic) wave, which induced an electron avalanche. Such clicks of avalanche detectors could be interpreted as the hit of photons, however, in the case under consideration we are dealing with a classical (although very weak) light wave, so such a photon interpretation is artificial and there is no need for it: discrete events recorded by the avalanche detector can be explained within the framework of classical electrodynamics or even classical wave optics [2-10].

Very weak classical light (electromagnetic) waves can be conditionally called "quantum" waves, which means only that they can be detected only by avalanche detectors and only in the form of discrete events (clicks), while the detailed characteristics (the amplitude of the electric field, phase and polarization) of such waves cannot be directly measured by classical macroscopic (i.e., consisting of a large number of atoms) devices. Note once again that when we speak of very weak classical light waves as "quantum" waves, we do not mean that they consist of discrete quanta - photons, but we only emphasize the discreteness of results of the interaction of such waves with avalanche detectors. From this point of view, we can leave the name "photon" as applied to a specific discrete event - the click of an avalanche detector induced by the action of a continuous classical light wave.

If a very weak classical light wave, regardless of its intensity, excites each time only one of the atoms of the avalanche detector which induces an electron avalanche, then the magnitude and the shape of the electric current pulse generated by the detector will not depend on the intensity of the incident wave. This means that we cannot determine the intensity of a very weak incident wave using the magnitude and shape of the electric current pulse; only the very fact of excitation of the avalanche detector can be detected. In turn, this means that a very weak light (electromagnetic) wave cannot be intercepted imperceptibly: even if we detect the click of the avalanche detector under the action of such a wave, we cannot to determine its parameters (intensity, electric field strength, phase, polarization) since the same actuation of the avalanche detector can occur under the action of a wave with other parameters. Moreover, as a result of the interaction of the incident wave with a detector, the incident wave will be absorbed (or, at least, significantly distorted) by the detector. This means that after the avalanche detector is triggered it will be impossible to create an exact copy of the absorbed wave and send it further, as if it was not intercepted.

Thus, we see that the detecting the very weak classical electromagnetic wave by a classical avalanche detector has all the properties of detecting photons: (i) amplification of the "quantum" properties of the wave (i.e., an increase in fluctuations) as the wave intensity decreases; (ii) the discreteness of the detected signal (which is can be interpreted as the photon hit on the detector); the appearance of a click of the detector during the interaction of a classical wave with a detector can be interpreted as a "collapse" of the wave; (iii) the disappearance of information about the incident wave after its absorption by the detector (which can be interpreted as the absorption of a "photon", loss of information about it and the impossibility of further manipulation of this "photon"); (iv) the inability to know the parameters (the amplitude of the electric field, phase, polarization) of a very weak light wave absorbed by the avalanche detector and, as a consequence, (v) the inability to create an exact copy of the absorbed wave (which can be interpreted as the inability to create an exact copy, i.e. to clone the absorbed "photon").

## 3 EPRB Gedankenexperiment with the weak classical light waves

EPRB Gedankenexperiment with classical light waves was analyzed in detail in [10]. Here, this experiment will be discussed in the volume necessary to understand the "quantum" key distribution protocol, which will be considered in the next section.

Let there is a source $S$ of a very weak radiation which emits two identical classical electromagnetic (light) waves in two opposite directions (Fig. 1):

$$\mathbf{E}_1 = \mathbf{E}_2 = \mathbf{E} \tag{6}$$

for light waves $v_1$ and $v_2$. As a result, both these waves will correlate, and this fact is obviously not surprising and does not lead to any paradoxes.

We assume that the radiation source $S$ is Gaussian, i.e. the components of the electric field vector of the light wave $\mathbf{E} = (E_x, E_y)$ are statistically independent and obey the normal distribution law

$$p(E_x, E_y) = \frac{1}{2\pi I_0} \exp\left(-\frac{E_x^2 + E_y^2}{2I_0}\right) \tag{7}$$

where

$$I_0 = \langle E_x^2 \rangle = \langle E_y^2 \rangle \tag{8}$$

is the characteristic of the source $S$; $\langle ... \rangle$ means averaging.

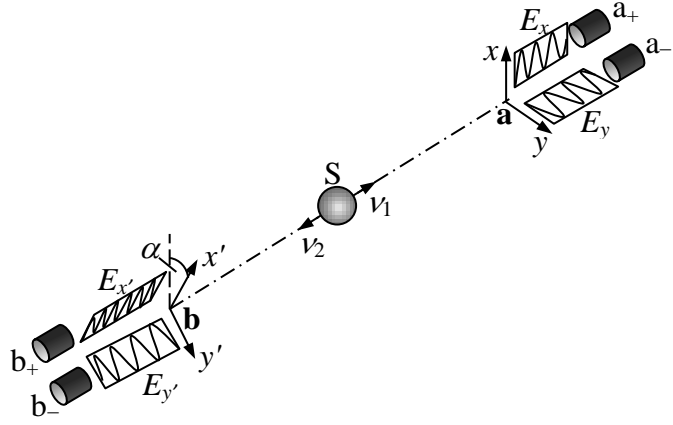

Fig. 1. Scheme of the "quantum" key distribution using the very weak classical light waves

One of these waves $\nu_1$ arrives at the polarizer **a** of Alice, which splits it into two mutually orthogonal linearly polarized beams, one of which (hereinafter referred to as the "+") has a polarization parallel to the $x$ axis of the polarizer, while the other (hereinafter referred to as the "-") has a polarization perpendicular to the axis of the polarizer. These beams arrive at the avalanche detectors, respectively, $a_+$ and $a_-$.

The second wave $\nu_2$ arrives at Bob's polarizer **b**, which is identical to the polarizer **a**, which splits it into two mutually orthogonal linearly polarized beams "+" and "-", arriving at avalanche detectors, respectively, $b_+$ and $b_-$.

The polarizers **a** and **b** can rotate about the axis of the incident beams.

All observation time is divided into discrete time windows of the same duration $\Delta t$, separated by identical time intervals. In each time window, Alice and Bob register which of their detectors triggered.

In general, the actuation of any detector in a given time window can either not occur, or occur one or more times. Therefore, the readings of the detectors can take the values $a_\pm = 0,1,2,...$ and $b_\pm = 0,1,2,...$, where the numbers indicate the number of triggering the corresponding detector in the given time window.

If we will try to interpret the results of the experiments from the point of view of the photonic representations, i.e. we assume that each click of the detector is a "photon" hit, then only those time windows are of interest for the "quantum" key distribution in which the conditions

$$a_+ + a_- = 1 \text{ and } b_+ + b_- = 1 \qquad (9)$$

are satisfied.

From the point of view of the photonic representations, this means that in this time window both Alice and Bob each detected exactly one indivisible "photon".

As shown in [10], the analysis of exactly such windows leads to the appearance of "entanglement" of weak classical light waves.

It is these windows hereinafter will be considered "good", and only they will be taken into account further when creating a "quantum" key.

Hereinafter, it is convenient to use nondimensional parameters [10].

Let us introduce the nondimensional exposure time (nondimensional duration of time window)

$$\tau = bI_0\Delta t \tag{10}$$

In this case, the probability of excitation of the atom during a time window is

$$P(t) = 1 - \exp\left(-(I/I_0)\tau\right) \tag{11}$$

Hereinafter, we take the value $I_0$ as a characteristic intensity of light. In this case we use the parameter $I_0^{1/2}$ as a scale for the field $\mathbf{E}$. Taking into account (7) and (11), in the subsequent calculations we will take $I_0 = 1$, while the value $I_0$ itself will be taken into account in the nondimensional duration of time window $\tau$.

Then the relations (7) and (11) can be written in nondimensional form:

$$P(\tau) = 1 - \exp(-I\tau) \tag{12}$$

$$p(E_x, E_y) = \frac{1}{2\pi}\exp\left(-\frac{E_x^2 + E_y^2}{2}\right) \tag{13}$$

where $I = \mathbf{E}^2$.

Let the total number of time windows equals to $N$. Leaving among these time windows only "good" ones, satisfying the condition (9), we can calculate the number of time windows in which simultaneously the pairs of corresponding events occurred: $N_{++}(\mathbf{a},\mathbf{b})$, $N_{--}(\mathbf{a},\mathbf{b})$, $N_{+-}(\mathbf{a},\mathbf{b})$, $N_{-+}(\mathbf{a},\mathbf{b})$. For example, $N_{++}(\mathbf{a},\mathbf{b})$ is the number of "good" time windows in which there were simultaneously detected one event on the detector $a_+$ and one event on the detector $b_+$, while at the same time, no events were observed on other detectors, etc.

The probabilities of the corresponding pair events (conditional probabilities) are determined by the relation

$$P_{\pm\pm}(\mathbf{a},\mathbf{b}) = N_{\pm\pm}(\mathbf{a},\mathbf{b})/N_0 \tag{14}$$

where

$$N_0 = N_{++}(\mathbf{a},\mathbf{b}) + N_{--}(\mathbf{a},\mathbf{b}) + N_{+-}(\mathbf{a},\mathbf{b}) + N_{-+}(\mathbf{a},\mathbf{b}) \tag{15}$$

is the total number of "good" time windows.

As shown in [10], it follows from (11) - (13) that

$$N_{++}(\mathbf{a},\mathbf{b}) = N_{--}(\mathbf{a},\mathbf{b}), \quad N_{-+}(\mathbf{a},\mathbf{b}) = N_{+-}(\mathbf{a},\mathbf{b}) \tag{16}$$

$$N_{++}(\mathbf{a},\mathbf{b})/N = \frac{1}{\sqrt{1+4\tau+4\tau^2\sin^2\alpha}} - \frac{2}{\sqrt{1+6\tau+8\tau^2}} + \frac{1}{(1+4\tau)} \tag{17}$$

$$N_{+-}(\mathbf{a},\mathbf{b})/N = \frac{1}{\sqrt{1+4\tau+4\tau^2\cos^2\alpha}} - \frac{2}{\sqrt{1+6\tau+8\tau^2}} + \frac{1}{(1+4\tau)} \tag{18}$$

where $\alpha$ is the - angle between the axes of polarizers $\mathbf{a}$ and $\mathbf{b}$.

Taking (16) into account, for conditional probabilities (14) one obtains the relation

$$P_{\pm\pm}(\mathbf{a},\mathbf{b}) = \frac{N_{\pm\pm}(\mathbf{a},\mathbf{b})}{2(N_{++}(\mathbf{a},\mathbf{b}) + N_{+-}(\mathbf{a},\mathbf{b}))} \tag{19}$$

Taking into account (19), one obtains

$$P_{++}(\mathbf{a},\mathbf{b}) = P_{--}(\mathbf{a},\mathbf{b}), \quad P_{-+}(\mathbf{a},\mathbf{b}) = P_{+-}(\mathbf{a},\mathbf{b}) \tag{20}$$

From (16) and (19) it follows that

$$P_{++}(\mathbf{a},\mathbf{b}) + P_{+-}(\mathbf{a},\mathbf{b}) = P_{+}(\mathbf{a}) = \frac{1}{2}, \quad P_{--}(\mathbf{a},\mathbf{b}) + P_{-+}(\mathbf{a},\mathbf{b}) = P_{-}(\mathbf{a}) = \frac{1}{2} \tag{21}$$

$$P_{++}(\mathbf{a},\mathbf{b}) + P_{-+}(\mathbf{a},\mathbf{b}) = P_{+}(\mathbf{b}) = \frac{1}{2}, \quad P_{--}(\mathbf{a},\mathbf{b}) + P_{+-}(\mathbf{a},\mathbf{b}) = P_{-}(\mathbf{b}) = \frac{1}{2} \tag{22}$$

where $P_{\pm}(\mathbf{a})$ and $P_{\pm}(\mathbf{b})$ are the probability of corresponding single events $a_{\pm} = 1$ and $b_{\pm} = 1$.

From the point of view of the "quantum" key distribution, we will only be interested in the case of the same orientation of the polarizers $\mathbf{a}$ and $\mathbf{b}$, i.e. when $\alpha = 0$.

In this case, the pointer of correlation of classical light waves $v_1$ and $v_2$ is the probability

$$P_{err} = P_{+-}(\alpha = 0) + P_{-+}(\alpha = 0) \tag{23}$$

that in "good" time windows the results of Alice and Bob do not coincide, i.e. the probability that in a "good" time window when the detector $a_+$ is triggered, the detector $b_-$ will trigger simultaneously or, vice versa, when the detector $a_-$ is triggered, the detector $b_+$ will trigger simultaneously.

Taking into account (20), one obtains

$$P_{err} = 2P_{+-}(\alpha = 0) \tag{24}$$

The dependence of the error probability $P_{err}$ on the nondimensional duration of the time window is shown in Fig. 2. For example, at $\tau = 20$ we obtain $P_{err} \approx 0.022$, that is, about in 2% of the "good" time windows, Alice and Bob's data will not coincide. This error is obviously commensurable or even less than the error that occurs in actual conditions due to the imperfection in the optical system and detectors, and also because of noise [11-13].

Let us consider what happens if the waves $\nu_1$ and $\nu_2$ are emitted by different incoherent sources with the same $I_0$. In this case, the intensities of waves $\nu_1$ and $\nu_2$ do not correlate with each other although they separately obey the distribution (7). Then, as is easy to show,

$$P_{++}(\mathbf{a},\mathbf{b}) = P_{--}(\mathbf{a},\mathbf{b}) = P_{+-}(\mathbf{a},\mathbf{b}) = P_{-+}(\mathbf{a},\mathbf{b}) = \frac{1}{4} \qquad (25)$$

for any angle $\alpha$.

In this case, taking into account the definition (23), we obtain the error probability $P_{err} = \frac{1}{2}$.

Thus, using the value of the error probability $P_{err}$, Alice and Bob can easily understand whether the waves detected by them were emitted by a single source $S$ or they were emitted by different uncorrelated sources, although with the same $I_0$.

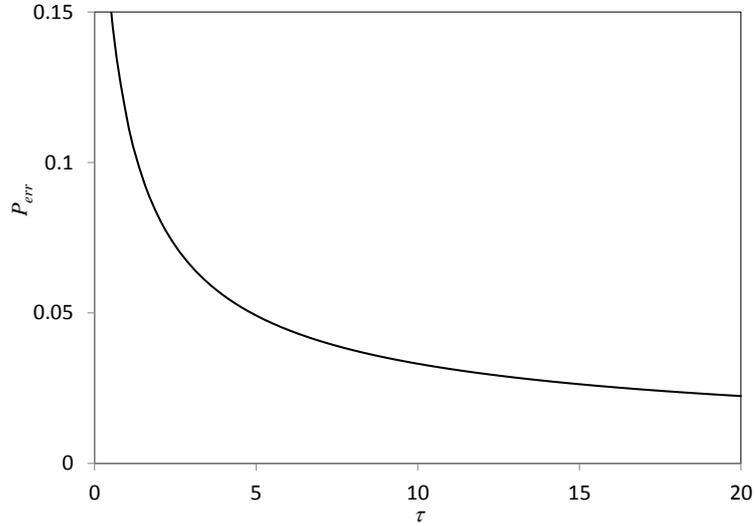

Fig. 2. Dependence of the error probability $P_{err}$ on the nondimensional duration of the time window $\tau$.

**4 "Quantum" key distribution protocol using the very weak classical light waves**

Suppose Alice wants to send a secret message to Bob. For this, she encrypts it using a key. The problem lies in the safe distribution of the key, taking into account that the third mandatory participant in this story, Eve, will try to intercept this key imperceptibly.

The secure protocol E91 for the key distribution, based on the use of "entangled photons", was proposed in [1] and was discussed from various perspectives in numerous works of followers.

The results obtained in the previous sections and in [10] allow constructing and implementing a secure key distribution protocol similar to the E91 protocol, but without photons, using very weak classical electromagnetic (light) waves and avalanche detectors.

Describing this "quantum" key distribution protocol, we will keep in mind the scheme shown in Fig. 1.

Alice and Bob agreed in advance on how the measurements will be made: what is the width of the time windows, during which a triggering the detectors will be recorded, what is the interval between the time windows, and also which of the rotation angles of the polarizers can be used. As usual, assume that only two admissible rotation angles of the polarizers $\theta = 0$ and $\theta = \pi/4$ (relative to some selected base coordinate system) are chosen.

Let us now formulate the "quantum" key distribution protocol.

1. Source $S$ (which controls, for example, Alice) sends to Alice and Bob weak light waves (6) - (8).

2. For each time window, Alice and Bob independently and randomly select one of the two permissible orientations of their polarizers and make measurements, recording which detectors triggered in each time window.

3. Bob selects "good" from his point of view time windows, i.e. windows that satisfy the condition $b_+ + b_- = 1$, and transmits their numbers to Alice via a public classical channel.

4. Alice selects from them "good" from her point of view time windows, i.e. those that satisfy the condition $a_+ + a_- = 1$, and transmits their numbers to Bob also via a public classical channel. After that, Alice and Bob have complete and identical sets of "good" time windows.

Now it is necessary to check whether the signals were intercepted by any third party (Eve).

5. To do this, Bob divides all the "good" time windows into two parts. Such a division can be carried out by prior agreement. Because all data are statistically homogeneous, then Bob can choose any part of the "good" time windows to check the correlation. For example, he can renumber all the "good" time windows and divide them into even and odd ones. Then Bob sends Alice via a public classical channel a complete information, for example, on even "good" time windows. Those, he informs Alice about the orientation of his polarizer and which detector triggered for these time windows.

6. Alice, using the information sent by Bob on even "good" time windows, checks them for correlation with hers corresponding time windows. To do this, she calculates the error probability (23).

7. If this probability is close to $\frac{1}{2}$ or at least substantially greater than the theoretical probability (Figure 2) for a given duration of time windows (i.e., even "good" time windows for Alice and Bob do not correlate or poorly correlated), this means that the signals were intercepted by Eve and cannot be used to create the key. Alice informs Bob about this, and this channel of communication is closes.

8. If Alice discovers that the even "good" time windows correlate in full accordance with the theory (that is, the calculated $P_{err}$ is close to the predictions of the theory with a correction on efficiency of the transceiver equipment), this means that the signals were not intercepted by a third party (Eve), and the remaining (odd) "good" time windows can be used to create the key. Alice inform Bob about this.

9. Bob, using a public classical channel, sends Alice an information about the orientations of the odd "good" time windows, but does not report the results of the detection in these windows.

10. Alice, using a public classical channel, sends Bob the numbers of those odd "good" time windows for which the orientation of her polarizer coincided with the orientation of Bob's polarizer, but also does not report the results of the detection. Now Alice and Bob have a complete set of odd "good" time windows in which the orientations of their polarizers coincided. The measurement results in these time windows are used by them to create the key. With probability 1-$P_{err}$, close to unity, Alice and Bob can be sure that in these time windows their results coincide. However, if Alice and Bob try to use their data directly to create the key, then their keys will be slightly different, and the probability of differences in their keys will be equal to $P_{err}$.

Taking into account the low probability of error, it can be easily corrected in various ways. The simplest way to correct this error is as follows. Because $P_{err} \ll 1$, then the number of errors in this key will be negligible. Therefore, Alice can encrypt with the help of her key any random, but meaningful message and send it via a public classical channel to Bob, who will try to decrypt it using his key.

When decoding, there will be some (small) number of errors, and Bob using the context of the message (for this purpose the message must necessarily have some sense) will be able to understand which of his key bits do not coincide with those of Alice. As a result, he simply changes the values of these bits to the opposite in his key.

After that, the secret key will be created and there will be complete confidence that it was not intercepted by a third party (Eve).

**5 Conclusions**

Thus, we have shown that "quantum" key distribution can be realized using very weak classical light waves and classical avalanche detectors in full accordance with the laws of classical physics. In doing so, we can realize all the features of the quantum E91 protocol.

It remains to answer the question: can Eve intercept imperceptibly a very weak classical light wave sent to Bob? Suppose Eve tries to divide the signal incoming to Bob into two equal parts: one of which she uses to intercept the key, and the other, after amplification, will be sent further to Bob so that he does not learn about the interception. In order to create the signals (intercepted by Eve and sent further to Bob) which do not differ from the signal sent by Alice, Eve should amplifies each of the obtained parts exactly by a factor of 2 and does not break their correlation with the intercepted signal, i.e. she must to keep the amplitude, phase and polarization unchanged. Only in this case, each of the parts of incident wave created by Eve will fully correlate with the wave received by Alice, and only in this case they will have the same effect on the avalanche detector as the original signal sent to Bob. As noted in Section 2, for a very weak light wave this cannot be done, because this would mean that we can measure the amplitude, phase and polarization of such a wave, and not just the fact of its detection (click of the detector). This means that, because we cannot in principle measure the parameters of a very weak light wave using of a macroscopic (consisting of a large number of atoms) measuring device (we can only record the click of an avalanche detector - the fact of its excitation by an incident wave), then we cannot amplify a weak electromagnetic wave by a predetermined number of times with the help of macroscopic devices, keeping its entanglement.

It was assumed above that the signals arriving at the polarizers in different time windows do not correlate. Let us explain what this means. From the mathematical point of view, the light wave (both $v_1$ and $v_2$) is a random process $\mathbf{E}(t)$ with zero mean $\langle \mathbf{E}(t) \rangle = 0$. This process is characterized by a two-time correlation function $R(T) = \dfrac{\langle \mathbf{E}(t)\mathbf{E}(t+T) \rangle}{\langle |\mathbf{E}(t)|^2 \rangle}$, such that $R(\infty) = 0$.

The value $T_0 = \int\limits_0^\infty R(T)dT$ is the characteristic decay time of the correlations (coherence time) and for light waves it is usually of the order of several nanoseconds (the characteristic lifetime of the excited state of the atom). Then the signals in neighboring (and, hence, in other) time windows can be considered uncorrelated (statistically independent) if the interval between time windows significantly exceeds $T_0$. This condition must be satisfied when implementing the "quantum" key distribution protocol described above.

Note that the process described in this article and in the article [10] (as, however, all the experiments on violation of Bell's inequality) is analogous to the widely discussed ghost imaging [14-17]. In fact, these are the same processes with the only difference that in the process under consideration a temporal correlation of the random light signal is used, while the ghost imaging uses the spatial correlation of the random field of the light wave. It is interesting that at present there are two competing interpretations of ghost imaging [14-17], one of which explains this process from the position of entangled photons, while the other considers the random classical correlating light fields.

**Acknowledgments** The work was supported by the Federal Agency for Scientific Organizations (State Registration Number AAAA-A17-117021310385-6). Funding was provided in part by the Tomsk State University competitiveness improvement program.